\documentclass[aps,prb,preprint,amsmath,amssymb,amsfonts,showkeys,showpacs,preprintnumbers]{revtex4-1}
\usepackage{bm,graphicx,xcolor,lineno,hyperref,microtype,multirow}

\newcommand{\eps}{\varepsilon}

\newcommand{\mr}{\multirow}

\newcommand{\ku}{k_{\uparrow}}
\newcommand{\kd}{k_{\downarrow}}
\newcommand{\upsa}{\Upsilon_0^{\text{a}}}
\newcommand{\upsb}{\Upsilon_0^{\text{b}}}
\DeclareMathOperator{\arccosh}{arccosh}

\begin{document}

\title{Exact energy of the spin-polarized two-dimensional electron gas at high density}

\author{Pierre-Fran\c{c}ois Loos}
\email{loos@rsc.anu.edu.au}
\affiliation{Research School of Chemistry, 
Australian National University, Canberra, ACT 0200, Australia}
\author{Peter M. W. Gill}
\thanks{Corresponding author}
\email{peter.gill@anu.edu.au}
\affiliation{Research School of Chemistry, 
Australian National University, Canberra, ACT 0200, Australia}
\date{\today}

\begin{abstract}
We derive the exact expansion, to $O(r_s)$, of the energy of the high-density spin-polarized two-dimensional uniform electron gas, where $r_s$ is the Seitz radius.
\end{abstract}

\maketitle

The three-dimensional uniform electron gas is a ubiquitous paradigm in solid-state physics \cite{Kohn99} and quantum chemistry, \cite{Pople99} and has been extensively used as a starting point in the development of exchange-correlation density functionals in the framework of density-functional theory. \cite{ParrYang}  The two-dimensional version of the electron gas has also been the object of extensive research \cite{Ando82, Abrahams01} because of its intimate connection to two-dimensional or quasi-two-dimensional materials, such as quantum dots. \cite{Alhassid00, Reimann02}

The two-dimensional gas (or 2-jellium) is characterized by a density $\rho = \rho_{\uparrow}+\rho_{\downarrow}$, where $\rho_{\uparrow}$ and $\rho_{\downarrow}$ are the (uniform) densities of the spin-up and spin-down electrons, respectively.  In order to guarantee its stability, the electrons are assumed to be embedded in a uniform background of positive charge. \cite{Vignale}  We will use atomic units throughout.

It is known from contributions by numerous workers \cite{Misawa65, Stern73, Zia73, Isihara77, Rajagopal77, Isihara80, Tanatar89, Attaccalite02, Seidl04, Giuliani07, Drummond09} that the high-density ({\em i.e.}~small-$r_s$) expansion of the energy per electron (or reduced energy) in 2-jellium is
\begin{equation} \label{Ejellium}
        E(r_s,\zeta) = \frac{\eps_{-2}(\zeta)}{r_s^2} + \frac{\eps_{-1}(\zeta)}{r_s} + \eps_0(\zeta) + \eps_\ell(\zeta) \,r_s \ln r_s + O(r_s),
\end{equation}
where $r_s = \left(\pi\rho\right)^{-1/2}$ is the Seitz radius, and 
\begin{equation}
	\zeta = \frac{\rho_{\uparrow} - \rho_{\downarrow}}{\rho}
\end{equation}
is the relative spin polarization. \cite{Vignale}  Without loss of generality, we assume $\rho_{\downarrow} \le \rho_{\uparrow}$, {\em i.e.} $\zeta \in [0,1]$.

The first two terms of the expansion \eqref{Ejellium} are the kinetic and exchange energies, and their sum gives the Hartree-Fock (HF) energy.  The paramagnetic ($\zeta=0$) coefficients are
\begin{align}
	\eps_{-2}(0) & = + \frac{1}{2},		\\
	\eps_{-1}(0) & = - \frac{4\sqrt{2}}{3\pi},
\end{align}
and their spin-scaling functions are
\begin{align}
	\Upsilon_{-2}(\zeta) & = \frac{\eps_{-2}(\zeta)}{\eps_{-2}(0)} = \frac{(1-\zeta)^{2}+(1+\zeta)^{2}}{2},		\label{ups-2-zeta}	\\
	\Upsilon_{-1}(\zeta) & = \frac{\eps_{-1}(\zeta)}{\eps_{-1}(0)} = \frac{(1-\zeta)^{3/2}+(1+\zeta)^{3/2}}{2}.	\label{ups-1-zeta}
\end{align}
In this Brief Report, we show that the next two terms, which dominate the expansion of the reduced correlation energy, \cite{Wigner32} can also be obtained in closed form for any value of the relative spin polarization $\zeta$. 

\newpage
The logarithmic coefficient $\eps_\ell(\zeta)$ can be obtained by a Gell-Mann--Brueckner resummation \cite{GellMann57} of the most divergent terms in the infinite series in \eqref{Ejellium}, and this yields \cite{Rajagopal77}
\begin{equation}
	\eps_\ell(\zeta) = - \frac{1}{12\sqrt{2}\pi} \int_{-\infty}^{\infty} \left[R\left(\frac{u}{\ku}\right)+R\left(\frac{u}{\kd}\right)\right]^3 du,
\end{equation}
where
\begin{equation}
\label{Ru-2D}
	R(u) = 1 - \frac{1}{\sqrt{1+1/u^2}},
\end{equation}
and
\begin{equation}
	k_{\uparrow,\downarrow} = \sqrt{1\pm\zeta}
\end{equation}
is the Fermi wave vector associated with the spin-up and spin-down electrons, respectively.  After an unsuccessful attempt by Zia, \cite{Zia73} the paramagnetic ($\zeta=0$) and ferromagnetic ($\zeta = 1$) values,
\begin{align}
	\eps_\ell(0) & = -\sqrt{2}\left(\frac{10}{3\pi}-1\right) = -0.0863136\ldots,	\\
	\eps_\ell(1) & = \frac{1}{4\sqrt{2}} \eps_\ell(0) = - \frac{1}{4}\left(\frac{10}{3\pi}-1\right) = -0.0152582\ldots,
\end{align}
were found by Rajagopal and Kimball \cite{Rajagopal77} and the spin-scaling function,
\begin{equation} \label{upsl-zeta}
	\Upsilon_\ell(\zeta) = \frac{\eps_\ell(\zeta)}{\eps_\ell(0)}
						= \frac{1}{8} \left[\ku+\kd+3 \frac{F\left(\ku,\kd\right)+F\left(\kd,\ku\right)}{10-3\pi}\right],
\end{equation}
was obtained 30 years later by Chesi and Giuliani. \cite{Giuliani07}  The explicit expression for $F(x,y)$ is
\begin{equation} \label{F}
	F(x,y) = 4(x+y) - \pi x - 4 x E\left(1-\frac{y^2}{x^2}\right) + 2x^2 \kappa(x,y),
\end{equation}
where
\begin{equation}
	\kappa(x,y) = 
		\begin{cases}
			(x^2-y^2)^{-1/2} \arccos(y/x),		&	x \le y,	\\
			(y^2-x^2)^{-1/2} \arccosh(x/y),		&	x > y,		\\
		\end{cases}
\end{equation}
and $E(x)$ is the complete elliptic integral of the second kind. \cite{NISTbook}

The constant coefficient $\eps_0(\zeta)$ can be written as the sum
\begin{equation} \label{eps0}
	\eps_0(\zeta) = \eps_{0}^{\text{a}}(\zeta) + \eps_{0}^{\text{b}}
\end{equation}
of a direct (``ring-diagram'') term $\eps_{0}^{\text{a}}(\zeta)$ and an exchange term $\eps_0^{\text{b}}$.  Following Onsager's work \cite{Onsager66} on the three-dimensional gas, the exchange term was found by Isihara and Ioriatti \cite{Isihara80} to be 
\begin{equation} \label{eps0-x-2D}
	\eps_{0}^\text{b} = \beta(2) - \frac{8}{\pi^2}\beta(4) = +0.114357\ldots,
\end{equation}
where $\beta$ is the Dirichlet beta function \cite{NISTbook} and $G=\beta(2)$ is Catalan's constant.  We note that $\eps_0^{\text{b}}$ is independent of $\zeta$ and the spin-scaling function therefore takes the trivial form
\begin{equation}
	\upsb(\zeta) = \frac{\eps_0^{\text{b}}(\zeta)}{\eps_0^{\text{b}}(0)} = 1.
\end{equation}

\begin{figure}
	\includegraphics[width=0.6\textwidth]{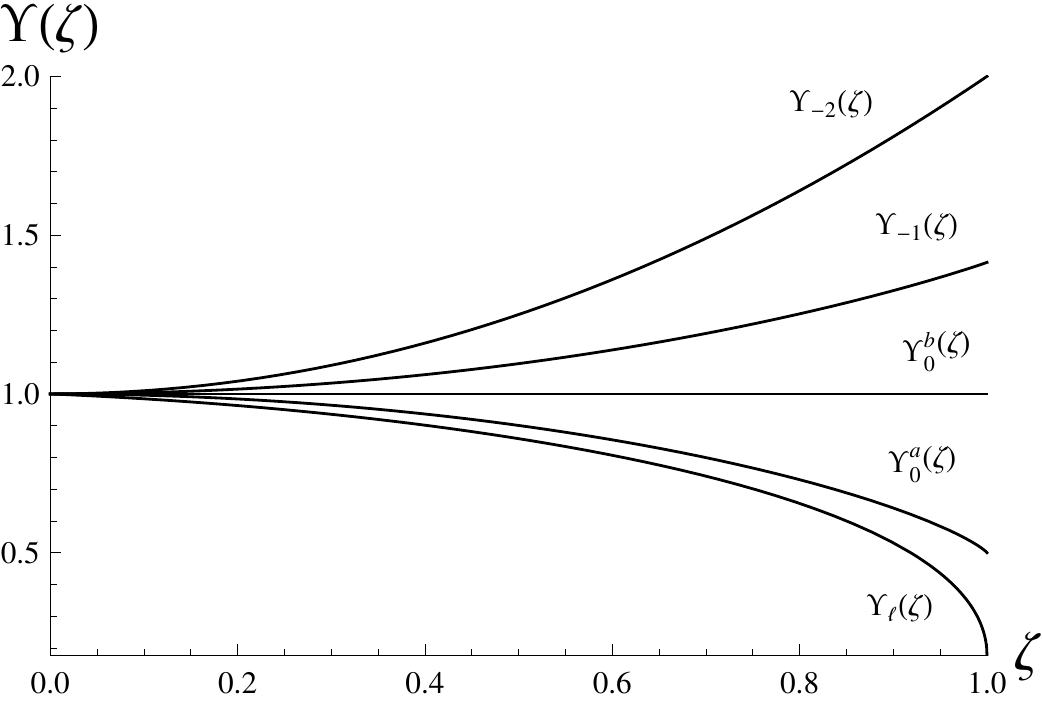}
	\caption{
	\label{fig:spin-scaling}
	$\Upsilon_{-2}(\zeta)$, $\Upsilon_{-1}(\zeta)$, $\upsa(\zeta)$, $\upsb(\zeta)$ and $\Upsilon_\ell(\zeta)$ as functions of $\zeta$.
	}
\end{figure}

The direct term has not been found in closed form, but we now show how this can be achieved.  Following Rajagopal and Kimball, \cite{Rajagopal77} we write the direct term as the double integral
\begin{equation} \label{eps0-d-int}
	\eps_{0}^{\text{a}}(\zeta) = -\frac{1}{8\pi^3} \int_{-\infty}^{\infty} \int_0^{\infty}
		\left[Q_{q/\ku}\left(\frac{u}{\ku}\right)+Q_{q/\kd}\left(\frac{u}{\kd}\right)\right]^2 dq\,du,
\end{equation}
where
\begin{equation}
	Q_{q}(u) = \frac{\pi}{q} \left[q-\sqrt{\left(\frac{q}{2}-i u-1\right) \left(\frac{q}{2}-i u+1\right)}
								-\sqrt{\left(\frac{q}{2}+i u-1\right) \left(\frac{q}{2}+i u+1\right)}\right].
\end{equation}
In the paramagnetic ($\zeta=0$) case, the transformation $s= q^2/4-u^2$ and $t=q\,u$ yields
\begin{equation}
	\eps_{0}^{\text{a}}(0) = -\frac{1}{2\pi} \int_{-\infty}^{\infty} \int_{0}^{\infty} \frac{1}{\sqrt{s^2+t^2}}
						\left[1-\left(\frac{\sqrt{(s-1)^2+t^2}+s-1}{\sqrt{s^2+t^2}+s}\right)^{1/2}\right]^2 dt\,ds,
\end{equation}
and, if we adopt polar coordinates, this becomes 
\begin{equation}
\begin{split}
	\eps_{0}^{\text{a}}(0)
		& = -\frac{1}{2\pi} \int_0^\infty \int_0^\pi
			\left[1-\sqrt{\frac{\sqrt{1-2r \cos\theta+r^2}-1+r\cos\theta}{r(1+\cos\theta)}}\right]^2 d\theta\,dr
		\\
		& = -\frac{1}{2\pi} \int_0^\pi
			\left[ 2\ln 2-(\pi-\theta) \tan\frac{\theta}{2}-2\tan^2\frac{\theta}{2} \ln\left(\sin\frac{\theta}{2}\right)\right] d\theta
		\\
		& = \ln 2 - 1
		\\
		& = -0.306853\ldots,
\end{split}
\end{equation}
which confirms Seidl's numerical estimate \cite{Seidl04} 
\begin{equation}
	\eps_{0}^{\text{a}}(0)=-0.30682\pm0.00012.
\end{equation}
In the ferromagnetic ($\zeta=1$) case, Eq.~\eqref{eps0-d-int} yields 
\begin{equation}
\label{eps0-d-1-2D}
	\eps_{0}^{\text{a}}(1) = \frac{1}{2} \eps_{0}^{\text{a}}(0) = \frac{\ln 2-1}{2} = -0.153426\ldots.
\end{equation}

In intermediate cases, where $0<\zeta<1$, we define the spin-scaling function 
\begin{equation}
	\upsa(\zeta) = \frac{\eps_{0}^{\text{a}}(\zeta)}{\eps_{0}^{\text{a}}(0)},
\end{equation}
and, from \eqref{eps0-d-int}, we have
\begin{equation}
	\upsa(\zeta) = \frac{1}{2} - \frac{1}{4\pi(\ln 2-1)} \int_{0}^{\infty} \int_{-1}^{1} P_{\ku}(r,z)  P_{\kd}(r,z) \frac{i\,dz}{z}\,dr,
\end{equation}
where
\begin{equation}
	P_{k}(r,z) = 1-\frac{\sqrt{r z-k^2} + \sqrt{r/z-k^2}}{\sqrt{r} \left(\sqrt{z}+1/\sqrt{z}\right)}.
\end{equation}
Integrating over $r$ gives
\begin{equation}
	\upsa(\zeta) = \frac{1}{2} - \frac{1}{4\pi(\ln 2-1)} \int_{-1}^{1} L_{\ku,\kd}(z) \frac{i\,dz}{z},
\end{equation}
where
\begin{equation}
\begin{split}
	L_{\ku,\kd}(z)	& = - \ku \ln \ku - \kd \ln \kd	
	\\
	& + \frac{1}{(z+1)^2}\Bigg[(z \ku - \kd)^2 \ln(z \ku - \kd) + (z \kd - \ku)^2 \ln(z \kd - \ku)	
	\\
	& -i \pi (\kd^2 - 2 z \ku\kd + \kd^2) + 2z (\ku+\kd)^2 \ln(\ku+\kd) - z(z\ku^2 - 2\ku\kd + z\kd^2) \ln z \Bigg],
\end{split}
\end{equation}
and contour integration over $z$ eventually yields
\begin{multline} 
\label{ups-zeta}
	\upsa(\zeta) 
	= \frac{1}{2} + \frac{1-\zeta}{4(\ln 2 - 1)} \Bigg[ 2\ln 2 - 1 - \sqrt{\frac{1+\zeta}{1-\zeta}}	\\
	+ \frac{1+\zeta}{1-\zeta}\ln\left(1 + \sqrt{\frac{1-\zeta}{1+\zeta}}\right) - \ln\left(1 + \sqrt{\frac{1+\zeta}{1-\zeta}}\right) \Bigg].
\end{multline}

This is plotted in Fig. \ref{fig:spin-scaling} and agrees well with Seidl's approximation, \cite{Seidl04} deviating by a maximum of $0.0005$ near $\zeta = 0.9815$.

In conclusion, we have shown that the energy of the high-density spin-polarized two-dimensional uniform electron gas can be found in closed form up to $O(r_s)$.  We believe that these new results, which are summarized in Table \ref{tab:coeffs}, will be useful in the future development of exchange-correlation functionals within density-functional theory.

We thank Prof. Stephen Taylor for helpful discussions.  P.M.W.G. thanks the NCI National Facility for a generous grant of supercomputer time and the Australian Research Council (Grants DP0984806 and DP1094170) for funding.

\begin{table*}
\caption{
\label{tab:coeffs}
Energy coefficients and spin-scaling functions for 2-jellium in the high-density limit.}
\begin{ruledtabular}
\begin{tabular}{ccccc}
	Term			&	Coefficient					&	$\eps(0)$	&	$\eps(1)$	&	$\Upsilon(\zeta)$		\\
	\hline																										\\
	$r_s^{-2}$		&	$\eps_{-2}(\zeta)$			&	$\displaystyle \frac{1}{2}$
													&	$\displaystyle 1$
													&	Eq.~\eqref{ups-2-zeta}									\\[10pt]
	$r_s^{-1}$		&	$\eps_{-1}(\zeta)$			&	$\displaystyle -\frac{4\sqrt{2}}{3\pi}$
													&	$\displaystyle -\frac{8}{3\pi}$
													&	Eq.~\eqref{ups-1-zeta}									\\[10pt]
	\mr{2}{*}{$r_s^0$}		&	$\eps_0^{\text{a}}(\zeta)$	&	$\displaystyle \ln2-1$
													&	$\displaystyle \frac{\ln 2 - 1}{2}$
													&	Eq.~\eqref{ups-zeta}									\\[10pt]
		&	$\eps_0^{\text{b}}(\zeta)$	&	$\displaystyle \beta(2)-\frac{8}{\pi^2}\beta(4)$
													&	$\displaystyle \beta(2)-\frac{8}{\pi^2}\beta(4)$
													&	1														\\[10pt]
	$r_s\ln r_s$	&	$\eps_\ell(\zeta)$			&	$\displaystyle -\sqrt{2}\left(\frac{10}{3\pi}-1\right)$	
													&	$\displaystyle -\frac{1}{4}\left(\frac{10}{3\pi}-1\right)$
													&	Eq.~\eqref{upsl-zeta}									\\[10pt]
\end{tabular}
\end{ruledtabular}
\end{table*}

\end{document}